\documentclass[12pt]{article}

\usepackage[utf8]{inputenc}
\usepackage{charter}
\usepackage{fullpage}
\usepackage{hyperref}
\usepackage{graphicx}
\usepackage{lipsum}
\usepackage[sort&compress,numbers]{natbib}
\usepackage{hyperref}
\hypersetup{hidelinks}
\usepackage[total={6.5in,9in},top=1in,headsep=0.1in,headheight=1in]{geometry}

\newcommand\snowmass{
\begin{center}
  \rule[-0.2in]{\hsize}{0.01in}\\
  \rule{\hsize}{0.01in}\\
  \vskip 0.1in
  Submitted to the Proceedings of the US Community Study\\ 
  on the Future of Particle Physics (Snowmass 2021)\\
  \rule{\hsize}{0.01in}\\
  \rule[+0.2in]{\hsize}{0.01in}\\[-2em]
\end{center}
}

\usepackage[firstpage=true]{background}
\backgroundsetup{contents={\parbox{6.5in}{\snowmass}}, scale=1,placement=top,opacity=1,color=black,position={3.25in,1.2in}}

\usepackage{fancyhdr}
\fancypagestyle{plain}{%
  \fancyhf{}%
  \fancyhead[C]{}
  \fancyfoot[C]{\thepage}
}

\fancypagestyle{empty}{%
  \fancyhf{}%
  \fancyhead[C]{{\it Snowmass2021 Cosmic Frontier White Paper Template}}
  \fancyfoot[C]{\thepage}
}
\title{Snowmass2021 Cosmic Frontier White Paper: High Density Galaxy Clustering in the Regime of Cosmic Acceleration}
\date{}

\usepackage{authblk}

\author[1]{Kyle Dawson}
\author[2]{Andrew Hearin}
\author[2]{Katrin Heitmann}
\author[3]{Mustapha Ishak} 
\author[4,5]{Johannes Ulf Lange}
\author[6,7]{Martin White}
\author[7]{Rongpu Zhou}
\author[\space]{\\ for the Snowmass 2021 Cosmic Frontier 4 Topical Group}
 
\affil[1]{Department of Physics and Astronomy, University of Utah,
Salt Lake City, UT 84112, USA}
\affil[2]{High Energy Physics Division, Argonne National Laboratory, 9700 South Cass Avenue, Lemont, IL 60439, USA}
\affil[3]{Department of Physics, The University of Texas at Dallas, Richardson, TX 75080, USA}
\affil[4]{Kavli Institute for Particle Astrophysics and Cosmology and Department of Physics, Stanford University, CA 94305, USA}
\affil[5]{Department of Astronomy and Astrophysics, University of
California, Santa Cruz, 1156 High Street, Santa Cruz, CA 95064, USA}
\affil[6]{Berkeley Center for Cosmological Physics, Department of Physics,
University of California, Berkeley, CA 94720, USA}
\affil[7]{Lawrence Berkeley National Laboratory, 1 Cyclotron Road, Berkeley, CA 94720, USA}

\begin{document}

\maketitle

\begin{abstract}

Joint studies of imaging and spectroscopic samples, informed by theory and simulations, offer the potential for comprehensive tests of the cosmological model over redshifts $z<1.5$. 
Spectroscopic galaxy samples at these redshifts can be increased beyond the planned DESI program by at least an order of magnitude, thus offering significantly more constraining power for these joint studies.
Spectroscopic observations of these galaxies in the latter half of the 2020's and beyond would leverage the theory and simulation effort in this regime. 
In turn, these high density observations will allow enhanced tests of dark energy, physics
beyond the standard model, and neutrino masses that will greatly exceed what is currently possible. 
Here, we present a coordinated program of simulations, theoretical modeling, and future spectroscopy that would enable precise cosmological studies in the accelerating epoch where the effects of dark energy are most apparent. 

\end{abstract}



\section{Introduction}\label{sec:intro}
Surveys to measure fluctuations in the cosmic microwave background (CMB) radiation and large-scale structure (LSS) through galaxies and quasars provide our best windows into the fundamental physics of the cosmos.
The tightest constraints on dark energy, modifications to General Relativity, mass limits on light dark matter particles, and neutrino masses all come from one or both of these measurements.  
Within the next few years, we expect to see a dramatic leap in our ability to probe fundamental physics through cosmology with the Dark Energy Spectroscopic Instrument \citep[DESI;][]{desi16a, desi16b}, Rubin Observatory's Legacy Survey of Space and Time (LSST), and CMB-S4.
DESI has already begun, and is quickly mapping the three dimensional distribution of galaxies and the intergalactic medium.  
The LSST on Rubin will begin in a few years, providing deep imaging over unprecedented areas of the sky.  
With CMB-S4 closing out the decade, our view of the cosmic microwave background will advance every bit as much as in the optical. 

With five years of spectroscopy from DESI, the expansion history of the universe will be well mapped.
However, growth of structure, the full shape of the power spectrum, higher order clustering statistics, and cosmology from joint lensing and spectroscopic surveys will be less well constrained.
The five-year DESI samples will allow initial investigations using these techniques, but advances in theory, simulations, and spectroscopic samples will allow significant advances.

In the latter half of this decade, DESI will remain the most competitive instrument in the world for large-scale spectroscopic surveys.
New spectroscopic facilities have been proposed with even faster survey speeds, offering the potential to increase the number density of DESI galaxies by an order of magnitude or more.
Long-term investments in spectroscopic surveys will allow enhanced tests of dark energy, physics beyond the standard model, and neutrino masses that will greatly exceed what is currently possible.  
A new sample at a much higher number number density that covers the full epoch of cosmic acceleration back to the matter dominated era will enable a unique spectroscopic probe of the following cosmological questions:

\begin{enumerate}
    \item Under General Relativity, structure grows at a rate that can be predicted at any time given the Hubble parameter, $H(z)$.
    New growth measurements over a range of redshifts will thus allow improved constraints on the expansion history and the equation of state for dark energy.
    \item By comparing the expansion history derived from growth measurements to that from geometric measurements, we can test the self-consistency of General Relativity or whether additional physics is required.
    \item The long redshift baseline afforded by growth measurements to $z<1.5$ will further break degeneracies of standard parameters with common dark energy models, and hence improve constraints on the dark energy equation of state.
    \item  The amplitude of clustering with redshift depends on the sum of neutrino mass.
    A survey of structure growth over a wide redshift range would tighten constraints on the neutrino masses.  If cosmology results favor the lowest allowed neutrino masses, then these measurements will also help resolve the question of mass hierarchy.
    \item Precise measurements of the power spectrum afforded by a large number of tracers would provide tight limits on dark matter interactions, light relic particles, early dark energy, and potential features in the primordial power spectrum that appear on small scales.
    \item Current lensing measurements of the amplitude of clustering indicate potential tension with predictions from the CMB under a $\Lambda$CDM model.
    As highlighted in the report from the Astro2020 Decadal Survey, resolving this tension with higher precision growth measurements could reinforce the $\Lambda$CDM model or reveal extensions in the dark sector.
\end{enumerate}

Order of magnitude increases over the initial DESI samples of $z<1.5$ galaxies are within reach of several proposed spectroscopic programs. 
A survey of galaxies at a high number density would allow high precision measurements down to small scales, with the statistical power to constrain the clustering amplitude at better than 0.1\% precision on scales comparable to a typical dark matter halo.
With a large sample of spectroscopically-confirmed galaxies, multiple tracers can be identified for clustering and cross-correlation studies, thus alleviating shot noise and offering robust checks on systematic errors.
Vastly increased sample sizes will also facilitate studies of higher order statistics that remain in the early stages of development.
Finally, by probing the same volume with spectroscopy and imaging data, such a survey will reap the full benefits of lensing and three-dimensional clustering.

In this white paper, we describe a program to use full-scale clustering (where both non-linear and linear scales are jointly modeled) to test the cosmological model over the redshift interval $0<z<1.5$.
In Section~\ref{sec:stage4}, we provide a brief overview of current Stage-IV Dark Energy programs and expectations for constraints on fundamental physics.
In Section~\ref{sec:analysis}, we describe the effort in theory and simulations that is required to reach the precision allowed by these new techniques.
Investment in that theory and simulation work should begin now for future spectroscopic observations to be most effective.
Finally, in Section~\ref{sec:facilities}, we describe the science drivers and new spectroscopic samples that are possible with existing and proposed facilities.

\section{Stage-IV Dark Energy Surveys}\label{sec:stage4}
\begin{table}[htb]
    \centering
    \begin{tabular}{|p{10cm}|p{5cm}|}
        \hline
        Question & Optimized approach \\ 
        \hline
        What is the physical origin of cosmic acceleration at late times (Dark energy)? & $z<1.5$ spectroscopy at high number density $+$
        Rubin lensing measurements \\ 
        \hline
        What is the sum and hierarchy of the neutrino masses?  & Full-shape power spectrum measurements from $z<1.5$ spectroscopy at high density $+$ characterization of underlying dark matter distribution through Rubin lensing measurements  \\
        \hline
        What is the particle nature of dark matter? & three-dimensional clustering on small scales with $z<1.5$ spectroscopy at high number density $+$ galaxy-galaxy lensing from Rubin, Roman, \& Euclid\\
        \hline 
        Does the rate of cosmic expansion and the growth of structure indicate new particle or field content? & All of the above \\
        \hline
    \end{tabular}
    \caption{P5 advances enabled by efforts at $z<1.5$ with theory, simulations, Rubin imaging, and DESI spectroscopy.  Spectroscopic sample sizes significantly larger than those planned with DESI are possible in the near future, allowing tighter constraints on each of these science drivers.}
    \label{tab:p5}
\end{table}

Large-scale cosmology surveys spanning imaging, spectroscopy, and lensing will independently advance four of the five key science drivers identified by the 2014 Particle Physics Project Prioritization Panel (P5) \cite{P5Report:2014pwa}.
Analyses that jointly utilize imaging and spectroscopy will substantially enhance the science returns that can be reaped from the data, as shown in Table \ref{tab:p5}. 

Stage-IV Dark Energy surveys will pursue these science drivers in three approximate redshift intervals using complementary techniques.  
At the lowest redshifts ($z<1$), Rubin will assess the distribution of dark matter on all scales through weak lensing, while DESI will assess the three dimensional clustering of galaxies with precise spectroscopy.  
Over moderate redshifts ($1<z<2$), Rubin will measure the distribution of galaxies through photometric redshifts, while DESI will provide three-dimensional positions with fiber-based spectroscopy.  
Finally, at the highest redshifts ($z>2$), Rubin will measure the distribution of galaxies through photometric redshifts, and DESI will map the three-dimensional distribution of matter through quasar and Lyman-$\alpha$ forest spectroscopy.  
In each of these redshift regimes, imaging and spectroscopy play critical, complementary roles.  DESI in particular will provide essential fiber-based, optical spectroscopy over the full redshift range $0<z<4$.

\subsection{DESI at $z<1.5$}

The currently planned DESI program will obtain a spectroscopic sample of roughly 13 million bright galaxies (BGS), seven million luminous red galaxies (LRG) at redshifts $z<1$, and 16 million emission line galaxies at redshifts $0.6<z<1.6$.
Covering roughly 2/3 of the observable extragalactic sky, the baryon acoustic oscillation (BAO) distance measurements from DESI over redshifts $z<1.0$ will be near the sample-variance limit.
Increasing the footprint or adding new $z<1$ targets will only have marginal returns for these BAO measurements.
The DESI BAO distance measurements over $1<z<1.5$ will be within a factor of two of the sample-variance limit for this same survey footprint, lending potential for future surveys of new areas and higher target densities.

These spectroscopic samples will also enable measurements of the growth of structure through redshift space distortions (RSD).
RSD result from the imprint of gravitational infall on the measured redshifts of galaxies.
These peculiar velocities record the history of structure growth which can be modelled with clustering measurements performed along and transverse to the line-of-sight.
RSD introduce anisotropy to this three-dimensional clustering, typically parameterized as $f \sigma_8 = \frac{\partial \sigma_8}{\partial\ln a}$,
where $a = (1 + z)^{-1}$ is the dimensionless cosmic expansion factor. 
Evaluating the amplitude of clustering at redshift $z=0$ under an assumption of a flat $\Lambda$CDM cosmology, $\sigma_8$ has been measured to a precision of 3.5\% \citep{eboss} using BAO and RSD in the Stage-III Baryon Oscillation Spectroscopic Survey \citep[BOSS;][]{dawson13} and its successor, eBOSS \citep{dawson16}.
Under the same $\Lambda$CDM model, with the same priors on the baryon density from Big Bang Nucleosynthesis (BBN), and the sample priors on the spectral index of the primordial power spectrum, the precision on $\sigma_8$ is expected to improve to 0.5\% in DESI when using similar algorithms and scales to $0.20\, \,h {\rm Mpc}^{-1}$ to perform the RSD measurements.
While a significant improvement over Stage-III, theoretical and modeling work will be needed to reach this goal.
Additional data will be needed to reach the precision of 0.2\% identified as a goal in the Astro2020 Decadal Survey to test consistency with the {\it Planck} \citep{planck20} $\sigma_8 = 0.811 \pm 0.006$ constraint.

\subsection{Rubin observatory at $z <  2$}

The Vera C. Rubin Legacy Survey of Space and Time (LSST) uses the Simonyi Survey Telescope, which is designed to observe a 18,000 deg$^2$ region of the sky in six optical bands~\cite{2019ApJ...873..111I}. LSST will run for a decade allowing the detection of $\sim$20 billion galaxies with photometric redshifts extending well-above unity. It will measure shapes for $\sim$2 billion galaxies  allowing lensing tomography to constrain the growth rate of structure and the equation of state of dark energy to reach the requirements for a Stage~IV dark energy experiments as defined in the Dark Energy Task Force report~\cite{2006astro.ph..9591A}. It will also measure clustering of galaxies and their cross correlation with lensing measurement in the so-called $3\times 2$pt analysis. LSST will also discover and measure at least 500 Supernovae Type~Ia (SNe~Ia) per season which will provide tens of thousands of well-measured SNe~Ia light curves up to $z\sim 1$ during the survey. LSST will discover $\sim$100,000 clusters of galaxies. Finally, LSST will measure strong gravitational lensing and time delay for the multiple images providing a sample of $\sim$2600 time-delayed lensing systems. A comprehensive analysis of the expected dark energy constraining power combining the probes can be found in Ref.~\cite{2018arXiv180901669T}.


\subsection{New Approaches with Stage-IV Dark Energy Surveys}

Whereas the expansion history will be well mapped by DESI at redshifts $z<1.5$, new techniques will be developed to improve clustering  measurements at smaller scales.
As a step toward this goal, several projects within DESI have begun to characterize DESI galaxies through galaxy-galaxy lensing from public weak lensing surveys (KiDS, DES, and HSC) and CMB lensing in order to deduce cosmological constraints from full-scale clustering studies.
Such cross-correlations often rely on a much higher number density than the BAO measurements, but may not need a spectrum of every object.  
Ongoing analyses will help establish the balance between photometric and spectroscopic sample sizes. 
Additional projects are also underway to compute higher order clustering statistics to better characterize scale-dependent bias that is degenerate with cosmological information at small scales.
These studies will also help establish the relationship between number density and cosmological constraints.

\section{Immediate Analysis Needs}\label{sec:analysis}
Theoretical forecasts of the cosmological constraining power of the nonlinear regime of structure formation now date back many years \citep[e.g.,][]{zentner_etal13,reid2014,krause_etal17,salcedo2020}, and suggest that an improved ability to use smaller-scale information can result in factors of 2-4 improvement on dark energy constraints. The natural question arises as to whether these gains can be realized in practice, or whether the need to marginalize over nuisance parameters capturing systematic uncertainty results in an excessive loss of constraining power. Recent work analyzing BOSS galaxy samples has shown that these potential gains could indeed be a reality \citep{wibking_etal20,lange_hearin_2021,chapman_etal21}; by including measurements from nonlinear scales, these recent analyses have achieved a full factor of 2 improvement beyond previous BOSS analyses \citep[e.g.][]{boss} that restricted attention to the quasi-linear regime. 

In addition to pushing the boundaries of a single survey by improving analysis capabilities, cross-correlations of data from different wavebands will enhance the science returns from ongoing and upcoming surveys (see Ref.~\cite{CrossCorr-Snowmass} for a detailed analysis).
Therefore, a modest investment into an overarching simulation and modeling program that will enable the exploitation of small scales and cross-correlations has the potential to substantially extend the scientific reach of ongoing and upcoming surveys (see also Ref.~\cite{2018arXiv180207216D}).

\subsection{High-Fidelity Simulations}

Simulations play an important role in fully exploiting the information available from cosmological surveys, in particular on smaller length scales. The simulation needs for ongoing and upcoming surveys span a wide range, from large numbers of realizations for covariance estimates, to detailed simulations that allow the creation of high-quality synthetic sky catalogs to enable tests of analysis pipelines and evaluation of systematic effects, to simulations that allow the exploration of physics beyond the confines of $\Lambda$CDM. Simulations beyond $\Lambda$CDM are important for many reasons. They provide predictions across cosmological parameters for a range of cosmological observables and enable the exploration of signatures of new physics beyond the current cosmological paradigm for both dark energy and dark matter.

The importance of simulation campaigns has been recognized widely in the community. Critical simulation campaigns have been carried out for contemporary surveys, including 
the MICE~\cite{2015MNRAS.448.2987F} and Buzzard simulations~\cite{2019arXiv190102401D} for DES, the DC2 simulated catalogs for LSST DESC~\cite{2019ApJS..245...26K,2021ApJS..253...31L}, and the AbacusSummit suite for DESI~\cite{2021MNRAS.508.4017M}. A related White Paper~\cite{CosComp-Snowmass} provides details regarding the roles of simulations in Cosmic Frontier science.

Conventional N-body simulations are now extremely mature tools; over the years, many code comparison and verification projects have been carried out, e.g., Refs.~\cite{2005ApJS..160...28H, heitmann2008cosmic, 2010ApJ...715..104H,2016JCAP...04..047S,osti_1501948,grove21}. These studies have shown that contemporary N-body codes with appropriate settings for initial conditions, force and mass resolution and coverage of large enough volumes, agree at the 1\% level out to scales of $k\sim 1 h^{-1}$Mpc, beyond which baryonic effects need to be taken into account. 
The statistical power of future spectroscopic surveys can be be 0.1\% or better at these scales, offering substantial power to constrain the assumptions in N-body simulations.
Simulation development over the course of the next decade is required to keep pace with the statistical precision allowed by these potential surveys.

\begin{figure}
    \centering
    \includegraphics[width=4.5in, trim=0 360 0 0, clip]{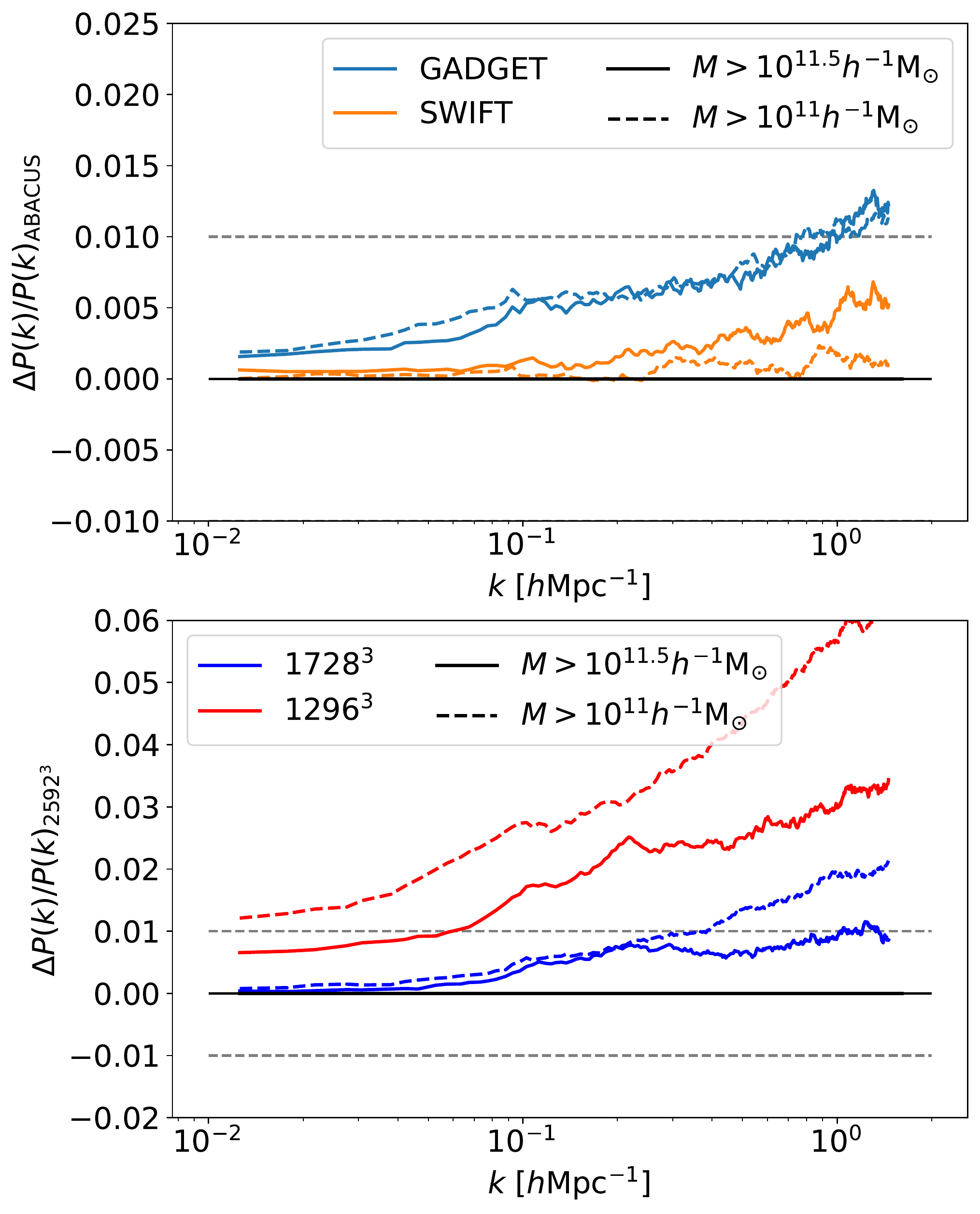}
    \caption{The real-space power spectra of halos from the N-body simulations using {\tt GADGET-2} \citep{springel05} and {\tt SWIFT} \citep{schaller16} relative to the {\tt ABACUS} \citep{garrison21} code.
    Figure reproduced from ``The DESI $N$-body Simulation Project I: Testing the Robustness of Simulations for the DESI Dark Time Survey'' \citep{grove21}. 
    }
    \label{fig:pksims}
\end{figure}

As a very recent example, in Ref.~\cite{grove21}, several N-body simulation codes have been considered within DESI to test theoretical models and systematic errors in two-point clustering and other measurement techniques. These simulations must accurately predict the growth of the dark matter halos that host the galaxies that are used as tracers of the underlying dark matter distribution. The simulations codes under consideration make different approximations to gravitational interactions between particles and therefore slightly different predictions for the halo mass distribution and its clustering properties. At scales out to $k_{\rm max}= 0.20 h\, \mathrm{Mpc}^{-1}$, where previous RSD measurements have been performed, there is almost perfect agreement between the high resolution {\tt SWIFT} simulations and the {\tt ABACUS} simulations (Figure~\ref{fig:pksims}). However, even for these two simulations, the agreement degrades to the 0.2\% level around $k = 1 h\, \mathrm{Mpc}^{-1}$. While these halos have not yet been populated with galaxies, the exercise demonstrates that a theoretical floor of at least 0.2\% precision currently exists in the first step of modeling halo clustering at small scales.

\subsection{Synthetic Skies from High-Fidelity Simulations}

Carrying out sufficiently accurate simulations is only the first step, however, establishing robust connections to the observable universe is essential for making useful predictions for galaxy clustering on nonlinear scales -- the process of galaxy formation is not directly modeled in N-body simulations. It is important, therefore, to develop improved techniques for relating the galaxy population to the underlying N-body simulations, a quantitative improvement in the so-called `galaxy-halo' connection.

Many contemporary efforts to derive cosmological constraints from nonlinear scales are built upon simplistic empirical models such as the Halo Occupation Distribution (HOD). Due to the very formulation of HOD-type models, incorporating new constraints from multiple redshifts and more than a single tracer galaxy population requires a significant expansion of the parameter space, and/or reliance upon plausibly-violated assumptions about the galaxy-halo connection. Thus conventional halo occupation models actually {\em penalize} attempts to incorporate new constraining data. Moreover, this older generation of models was devised at a time when consensus in the field had not yet been reached on the reliability of cosmological simulations to resolve halo substructure, and so the HOD and related models are founded merely upon host halos identified at a particular simulated snapshot. But the field of computational cosmology has seen dramatic progress in the quality of simulated data products over the last fifteen years: subhalo catalogs with merger trees are becoming widely available for high-resolution, survey-scale simulations \cite{Chuang_etal19_unit_sims,heitmann_etal_last_journey,Ishiyama_etal21_uchuu,bose_etal21}; thus the continued use of HOD-type models as the basis of our cosmological predictions fails to capitalize upon the now well-established ability of contemporary codes to track the evolution of halos and their substructure across cosmic time. These core limitations of traditional halo occupation models highlight how they bear the mark of the era in which they were developed. Hydrodynamical simulations or Semi-Analytic Models (SAMs) have also been used to generate predictions across redshift in a physically motivated way. These models remain irreplaceable in the effort to understand the detailed physics of galaxies and clusters, but the full scientific potential of data in the 2020s can only be delivered with expansive explorations of parameter space based on high-resolution, Gpc-scale simulations, and so new techniques beyond the traditional implementations of these models are needed.

Considerable recent progress has been made by a promising new generation of empirical models that bridge the gap between the level of complexity achieved by SAMs and the computational efficiency of empirical models, e.g., UniverseMachine \citep[][]{behroozi_etal18} and EMERGE \citep[][]{moster_etal17}. The ability of these models to make predictions for multiple tracers across redshift is promising, but further advances are needed on both the modeling and computation side for this new approach to conduct cosmological inference with survey-scale simulations. To meet the predictive needs and to maximize the scientific returns of the upcoming surveys, we consider it critical to invest in the development of a new generation of modeling approaches that address the limitations of contemporary techniques. 


\subsection{Lack of Inter-collaboration structures}

The general requirements on simulations to analyze surveys are similar for the different surveys. While some surveys need larger volumes, other surveys possibly require higher mass resolution. However, developing an overarching simulation program that allows surveys to share simulations and derived data products such as synthetic catalogs would be very valuable. In Ref.~\cite{2020arXiv200507281B} a strong case was made for a joint simulation program across the Vera Rubin Observatory (Rubin) Dark Energy Science Collaboration (DESC), the Nancy Grace Roman Space Telescope, and Euclid.
This case clearly carries over to DESI, CMB-S4, and other future surveys as well. 

\section{Future Survey Prospects}\label{sec:facilities}

Advances in simulations and the theory connecting galaxies to dark matter will allow more accurate modeling of the cosmological signal.
Meanwhile, future spectroscopic programs have the potential to significantly improve the statistical precision allowed in these analyses.
After DESI completes, there will still be a very large number of spectroscopic modes available to probe clustering at small scales at $z<1.5$.
New spectroscopic galaxy samples at these redshifts will allow precise measurements of structure growth and the full-shape of the power spectrum during the critical time when the Universe transitioned from a matter-dominated expansion to a dark energy-dominated expansion. 

\begin{figure}[t]
    \centering
    \includegraphics[width=6.5in]{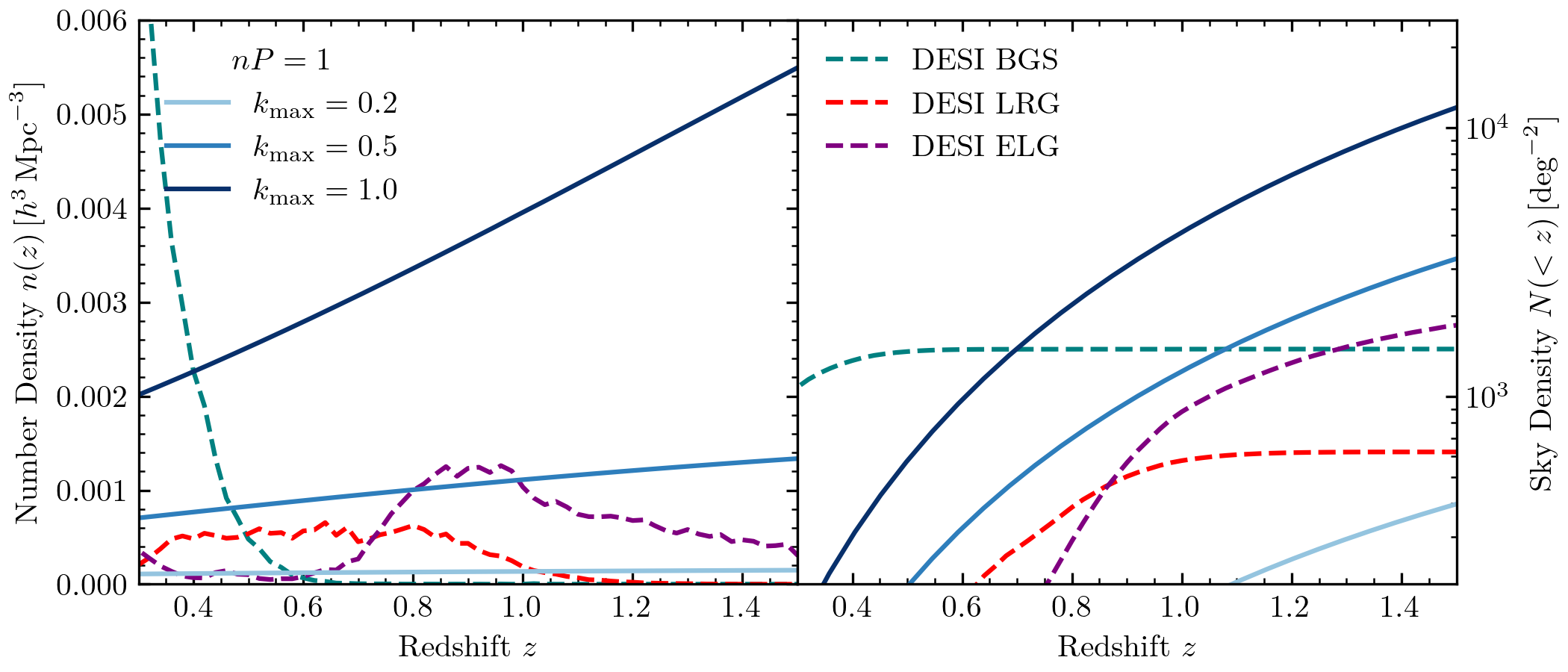}
    \caption{Number of spectroscopic tracers required to reach $\bar n P(k) = 1$ as a function of redshift for scales $k_{\rm{max}} = $0.2, 0.5, and 1.0 $h \, \rm{Mpc}^{-1}$.
    The DESI BGS, LRG, and ELG number densities are as expected based on current survey operations.
    {\bf Left:}  Number density as a function redshift.
    {\bf Right:}  Cumulative surface density as a function of redshift.
    The dependence of large-scale bias on number density is based on extrapolations of the HOD models \citep{zheng07} derived from clustering in the photometric catalogs of the DESI LRG sample \citep{zhou21}, an assumed halo mass function \citep{despali16}, and an assumed halo mass-halo bias relation \citep{tinker10}.
    The large-scale bias decreases by a factor of roughly 1.7 between the number densities reported for $k_{\rm{max}} = $0.2 $h \, \rm{Mpc}^{-1}$ and those reported for $k_{\rm{max}} = $1.0 $h \, \rm{Mpc}^{-1}$.
    }
    \label{fig:np1}
\end{figure}

Number density, redshift range, spectroscopic completeness, and volume for a future spectroscopic program will need to be matched to analysis techniques and a systematic error budget that will steadily improve over the next decade.
In the absence of a specific systematic error budget or well-accepted metrics to optimize area and number density for cross-correlation studies or higher order statistics, we refer to the product $\bar n P(k) = 1$ for characterizing the potential of future samples.
Here, $\bar n$ is the average number density of the sample and $P(k)$ is the amplitude of the observed power spectrum at a scale $k$.
This parameterization reflects a fairly optimal balance between area and number density for two-point statistics in a sample of a fixed size \citep[for further discussion, see][]{forecasts14}.
As shown in the left panel of Figure~\ref{fig:np1}, the DESI luminous red galaxy (LRG) and emission line galaxy (ELG) samples will achieve number densities that exceed the limit $\bar n P(k) = 1$ for scales $k_{\rm{max}} = 0.20 \, h \, \rm{Mpc}^{-1}$.

The bright galaxy sample (BGS) will reach $\bar n P(k) = 1$ for scales $k_{\rm{max}} = 1 \, h \, \rm{Mpc}^{-1}$ at redshifts $z<0.3$.
A scale of $k_{\rm{max}} = 1 \, h \, \rm{Mpc}^{-1}$ corresponds roughly to the transition between sampling the halo population (two halo term) and sampling the galaxies within a halo (one halo term).
It is this region where the nuisance terms resulting from connecting galaxies to halos are best constrained, thus breaking degeneracies with the cosmological information.

The DESI BGS sample has such a high number density in part to serve as a testbed for new cosmological analysis techniques.
Future programs have the potential to expand these techniques to higher redshifts and larger volumes.
As shown in the right panel of Figure~\ref{fig:np1}, a program that achieves $\bar n P(k) = 1$ at $k_{\rm{max}} = 1 \, h \, \rm{Mpc}^{-1}$ to a redshift $z=1$ would require roughly 4000 galaxies deg$^{-2}$.
A program that extends a clustering sample of this scope to $z=1.5$ would require roughly 10,000 galaxies deg$^{-2}$.
With samples of this size extending to higher redshift, the techniques developed on the BGS sample can eventually be applied over a redshift range that includes the matter dominated era and the transition to the accelerating epoch.


\subsection{Science Drivers for Low-z clustering with future surveys}
Current lensing measurements indicate tension with predictions from the CMB under a $\Lambda$CDM model (see, e.g., Ref.~\cite{2021A&A...646A.140H} for recent results).
These tensions may be explained by an additional field or interaction that distorts the shape of the power spectrum, modifications to General Relativity, other extensions to the Standard Model, a chance statistical realization of the analyses, or systematic errors in the data analysis.
Three-dimensional clustering data from DESI and lensing from Rubin observatory will allow us to improve the precision of these measurements.
Advances in simulations will allow us to improve the theoretical modeling of these measurements.
New spectroscopic data samples that sample clustering at the densities presented in Figure~\ref{fig:np1} over a wide redshift range can advance our understanding of cosmic acceleration as follows:

\begin{enumerate}
\item{Growth of structure:  probe origin of cosmic acceleration at late times}

\item{Growth of structure:  constrain neutrino mass}

\item{Full-shape power spectrum measurements:  provide limits on models for new physics that distort the shape of the power spectrum at scales $k < 1 \, h \, \rm{Mpc}^{-1}$ }

\item{Calibrating samples of voids, clusters, peculiar velocities, and other techniques:  perform comprehensive tests of the cosmologial model}

\item{Higher-order statistics:  resolving degeneracies or identifying cosmological signatures not present in the power spectrum}

\item{Multiple probes:  test for new particle or field contents in the cosmological model}
\end{enumerate}

The techniques to perform these cosmological constraints will likely arise from a combination of three-dimensional clustering at small scales, higher order clustering statistics, and cross-correlations with lensing surveys.

\subsection{Clustering Samples from DESI with Extended LRG Selections}\label{sec:LRG}
\begin{figure*}[t]
\centering
\includegraphics[width=3.2in]{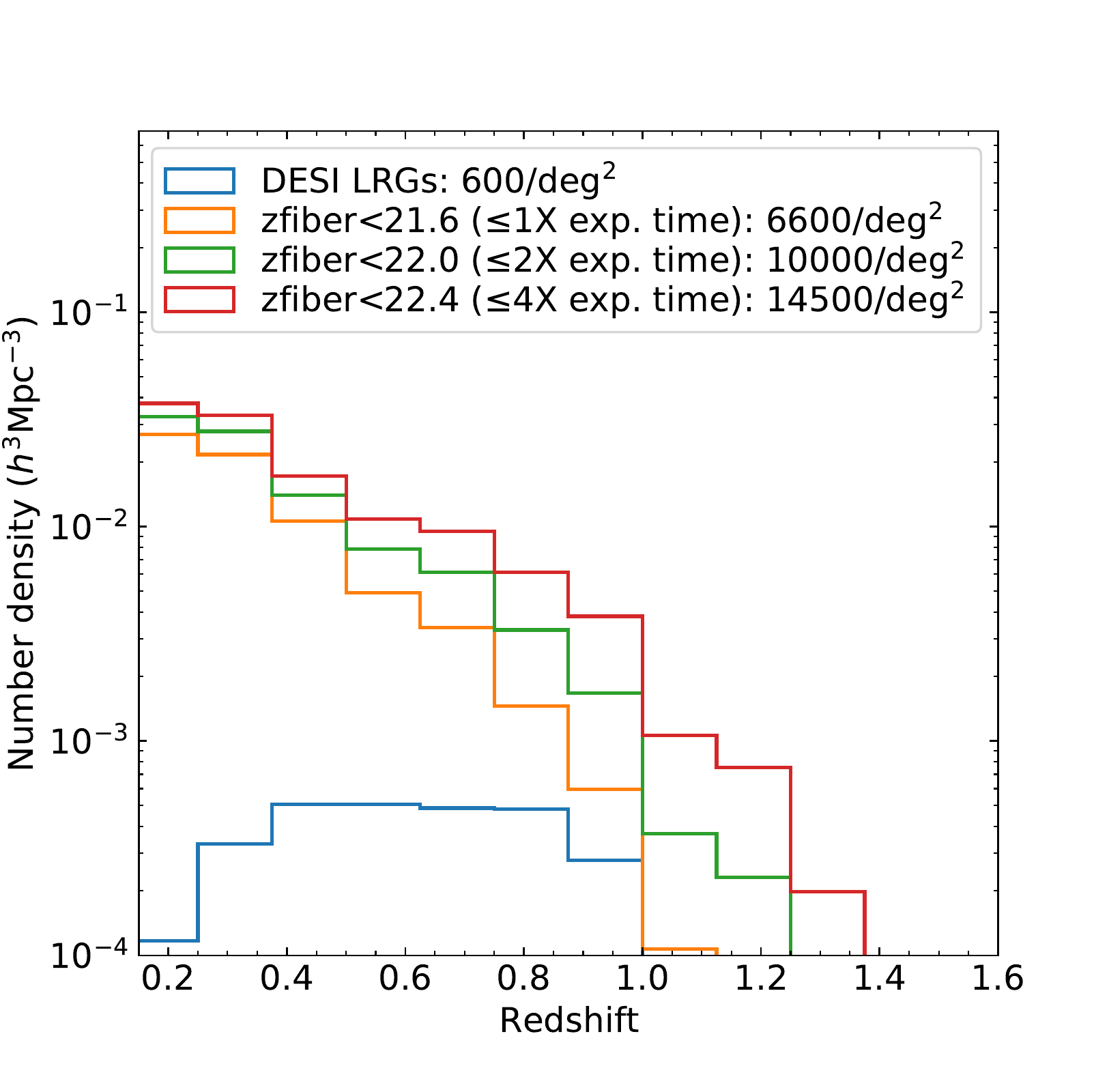}
\includegraphics[width=3.2in]{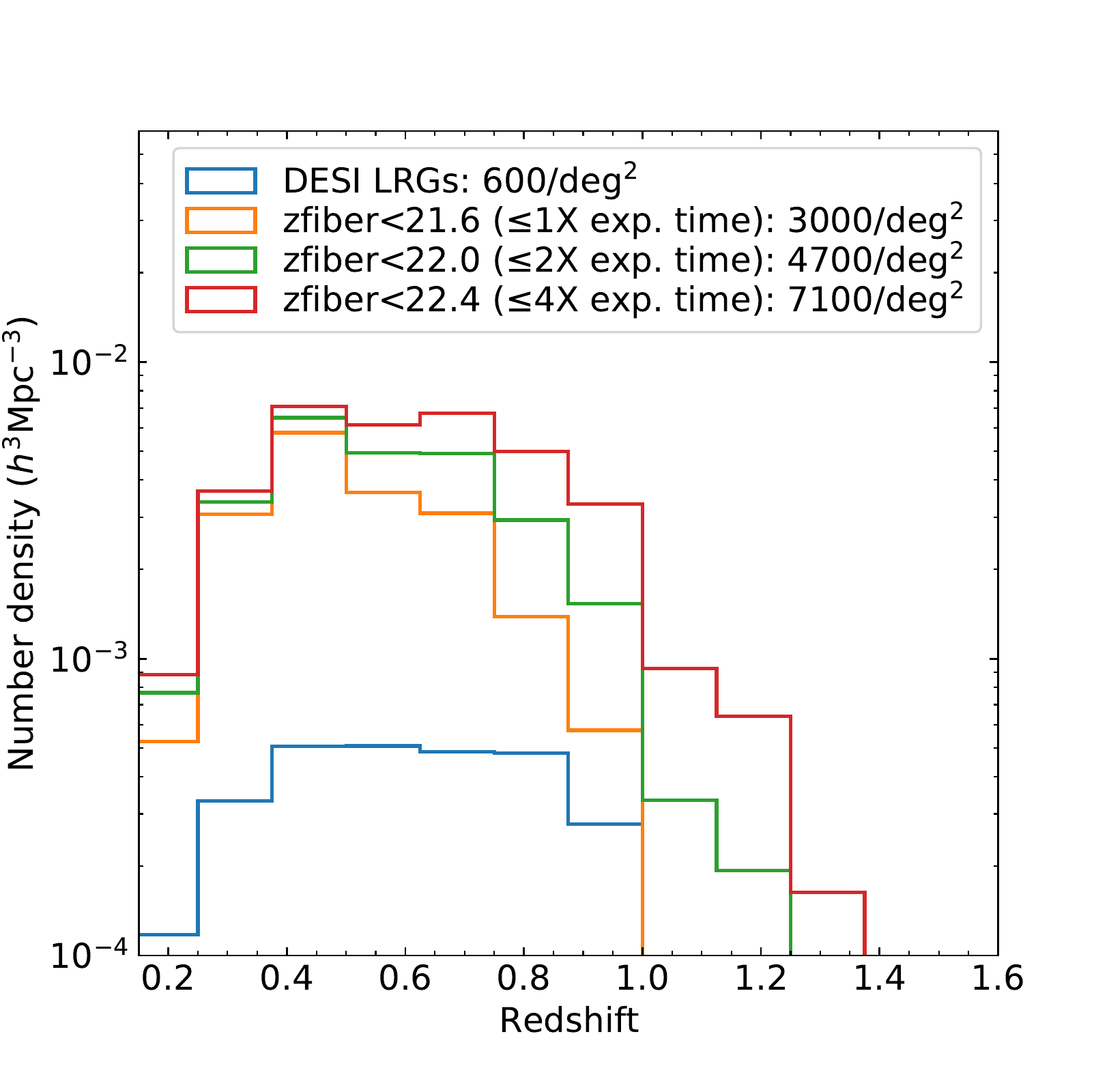}
\caption{Expected redshift distribution for the DESI LRG samples compared to various extended selection algorithms.  {\bf Left:  }A broad extension to the DESI algorithm that excludes criteria intended to reduce the number of low redshift galaxies.  {\bf Right:  }An extended selection that incorporates criteria intended to reduce the number of $z<0.3$ galaxies.
In both cases, the redshifts are from COSMOS photo-z's \citep{laigle_cosmos2015_2016} for the $z_{\rm fiber}$-limited samples and DESI spectroscopic redshifts for DESI LRGs. Pixel-level simulations indicate that the samples described by a selection to $z_{\rm fiber}<21.6$ can be spectroscopically classified in the same exposure time as planned for the ongoing DESI program.
Simulations indicate that galaxies with $21.6<z_{\rm fiber}<22.0$ require exposures twice as long while galaxies with $22.0<z_{\rm fiber}<22.4$ require exposures four times as long.
}
\label{fig:LRG}
\vspace{-0.2cm}
\end{figure*} 

The reddest, most luminous LRGs exhibit relatively high large-scale-structure bias, which enhances the amplitude of their clustering.
Because of their strong 4000 \AA\ break and their well-behaved, red spectral energy distributions, LRGs
can be selected from imaging catalogs and spectrally classified with high efficiency.
Pixel-level simulations indicate that more than 95\% of these targets can be assigned reliable redshifts using a selection to $z_{\rm fiber}<21.6$, where $z_{\rm fiber}$ is the $z$-band magnitude in an aperture matched to the DESI fiber diameter.

The currently planned DESI program will obtain an LRG spectroscopic sample of roughly six million galaxies at redshifts $z<1$.
Beyond this approximate redshift, the strong 4000 \AA\ break and characteristic spectral features move into a region of the spectrum with strong sky line contamination, thus complicating redshift classification.
As shown in Figure~\ref{fig:LRG}, the DESI LRG sample will cover the redshift range $0.3<z<1$ with a surface density of approximately 600 deg$^{-2}$.
These observations will be obtained over five years, using a subset of the 50M fibers that will be available over all pointing centers.

In the latter half of this decade, DESI will remain the most competitive instrument for large-scale spectroscopic surveys.  
DESI will be able to maintain a high redshift efficiency for targets identified in extensions to the LRG selections that include bluer galaxies with lower bias.
Potential selections have been explored using infrared and optical photometry, with
simulations revealing consistently high redshift efficiency at the same exposure times planned for the current DESI survey.
Two such selections are presented in orange in Figure~\ref{fig:LRG}.
At 6600 deg$^{-2}$, the first selection (left panel) includes a high fraction of low redshift galaxies and very high completeness as a function of stellar mass.
The second selection (right panel) is tuned for minimal redshift overlap with the BGS sample, leading to a density of 3000 deg$^{-2}$.
This second selection exceeds a number density of $1 \times 10^{-3} \, h^3 \, Mpc^{-3}$ out to $z<0.9$.
As demonstrated in Figure~\ref{fig:np1}, this number density would be sufficient to reach the threshold $\bar n P(k) = 1$ at $k_{\rm{max}} = 1 \, h \, \rm{Mpc}^{-1}$.
The galaxies in these two selections are sufficiently bright that spectra can be classified at high efficiency using the same exposure times as planned for the current LRG sample.
At this exposure time, roughly 10M spectra can be obtained per year.

Longer exposure times could be used to include even fainter galaxies, as shown in the green and red curves of Figure~\ref{fig:LRG}.
Extending the magnitude limit to $z_{\rm fiber}<22.0$ (green) would require a factor of two increase in exposure time for the additional 1700 deg$^{-2}$ galaxies included in the selection.
Doing so would extend the redshift coverage to $z<1$ where the density allows sampling of clustering beyond the shotnoise limit.
A factor of four increase in exposure time (red) would allow observations of a sample to $z_{\rm fiber}<22.4$, thus vastly exceeding the required number densities over $0.3<z<1$ and extending coverage to $z<1.1$.

Continuing DESI observations in the area that overlaps the Vera Rubin Observatory survey footprint would allow joint studies of a high density spectroscopic sample and high signal-to-noise lensing sample.
In Table~\ref{tab:DESI-II}, we demonstrate several potential survey strategies that can be completed with the DESI instrument in five years or less.
With only an investment in DESI operations, such a survey would allow enhanced tests of dark energy, physics beyond the standard model, and neutrino masses using very high density probes of large-scale structure out to an approximate redshift of $z=1$.  

\begin{table}[htb]
    \centering
    \begin{tabular}{c c c c}
        \hline
        Survey Boundaries & Area & Survey Time (years) & Sample Size \\ 
        \hline
        \hline
\multicolumn{4}{c}{Selection at 6600 deg$^{-2}$ without low redshift cut ($z_{\rm fiber}<21.6$)}\\
        \hline
$0<\delta < 12$ & 3372 deg$^2$ & 2 years &  22M \\
$-10<\delta < 12$ & 6024 deg$^2$ & 4 years &  40M \\
        \hline
\multicolumn{4}{c}{Selection at 3000 deg$^{-2}$ with low redshift cut ($z_{\rm fiber}<21.6$)}\\
\hline
$0<\delta < 12$ & 3372 deg$^2$ & 1 year &  10M \\
$-10<\delta < 12$ & 6024 deg$^2$ & 2 years &  18M \\
        \hline
\multicolumn{4}{c}{Selection at 4700 deg$^{-2}$ with low redshift cut ($z_{\rm fiber}<22.0$)}\\
        \hline
$0<\delta < 12$ & 3372 deg$^2$ & 2 years &  16M \\
$-10<\delta < 12$ & 6024 deg$^2$ & 4 years &  28M \\
        \hline
\multicolumn{4}{c}{Selection at 7100 deg$^{-2}$ with low redshift cut ($z_{\rm fiber}<22.4$)}\\
        \hline
$0<\delta < 12$ & 3372 deg$^2$ & 5 years &  24M \\
        \hline
    \end{tabular}
    \caption{Potential survey footprints for four different selection algorithms using DESI as shown in Figure~\ref{fig:LRG}.  The reported area corresponds to the overlap with Vera Rubin imaging in regions at galactic latitudes that avoid excessive Galactic extinction and stellar contamination.}
    \label{tab:DESI-II}
\end{table}

\subsection{Future Clustering Samples at $1<z<1.5$}

As shown in Figure~\ref{fig:np1}, roughly 6000 galaxies deg$^{-2}$ are required to reach a threshold of $\bar n P(k) = 1$ at $k_{\rm{max}} = 1 \, h \, \rm{Mpc}^{-1}$ over the redshift range $1<z<1.5$.
At these redshifts, the absorption features that allow robust classification of passive, LRG spectra become harder to detect due to increased sky background and decreased continuum signal.
For this reason, the DESI survey relies on the detection of [O~\textsc{ii}] line flux for galaxy redshift classification in this redshift range.
At a wavelength of 3727 \AA, the [O~\textsc{ii}] emission line that is often associated with star formation is visible with Silicon detectors to approximately $z<1.6$.
However, selecting these targets from imaging data to have both significant line strength and redshifts in the range $1.0<z<1.5$ is far more challenging than the selection of LRG targets described in Section~\ref{sec:LRG}.
As can be seen in the right panel of Figure~\ref{fig:np1}, the DESI ELG selection produces only around 400 deg$^{-2}$ robust redshift classifications in this redshift range.
These redshifts are obtained from a spectroscopic sample of roughly 1400 deg$^{-2}$, some of which are at lower redshift and some of which are not classified with high confidence.
More than an order of magnitude increase in number density is required to reach the shot-noise limited densities over this redshift range.

If a selection algorithm with new imaging data is identified to select strong [O~\textsc{ii}] emitting galaxies over $1<z<1.5$ with high efficiency, then it will be possible to improve clustering statistics with new DESI observations.
However, it is most likely that new selections will experience similar contamination from low redshift galaxies and require longer exposure times to reach to fainter [O~\textsc{ii}] line fluxes.
While such an investigation remains to be done, one can assume similar rates of redshift success.
The [O~\textsc{ii}] luminosity distribution is also unknown, but for illustrative purposes, we assume that an order of magnitude increase in sample size will produce limiting [O~\textsc{ii}] line strengths that are between a factor of one and two smaller than in the DESI sample. 
Such a sample would require between 21,000 and 84,000 fiber-exposures deg$^{-2}$ at equivalent DESI exposure times.
A pilot survey with DESI could complete between 120 and 500 deg$^2$ in one year under these assumptions.

Observations at this high density are more efficiently pursued with a future spectroscopic facility with a significantly higher survey speed.
Three concepts for future spectroscopic programs were described in white papers submitted to the Astro2020 Decadal Survey:  MegaMapper~\cite{2019BAAS...51g.229S}, Mauna Kea Spectroscopic Explorer (MSE)~\cite{2019BAAS...51g.126M}, and SpecTel~\cite{2019BAAS...51g..45E}.
Each of these offers the potential for even larger sample sizes than would be obtained with DESI.
Both the proposed fiber positioner designs and the collecting areas are yet to be finalized for these three facilities.
Accepting these uncertainties, we approximate the collecting area of the 6.5-meter MegaMapper telescope to be three times that of DESI and the collecting area of both the 12-meter class MSE and SpecTel telescopes to be ten times that of DESI.
We assume a similar focal plane of 20,000 fibers for all three designs, and assume that sufficient targets are available to populate the fibers regardless of field of view.
The survey areas available to each facility simply scales with the product of fiber number and collecting area.

Under these idealized assumptions, we find that in order to reach $\bar n P(k) = 1$ at $k_{\rm{max}} = 1 \, h \, \rm{Mpc}^{-1}$ over the redshift range $1<z<1.5$:

\begin{itemize}
    \item MegaMapper could complete between 1,500 and 6,000 deg$^2$ per year;
    \item MSE could complete at least 5000 deg$^2$ per year; and
    \item SpecTel could complete at least 5000 deg$^2$ per year.
\end{itemize}


\bibliographystyle{unsrt}
\bibliography{main.bib}

\end{document}